\documentstyle[psfig,aps,prl]{revtex}

\def\scr#1{\mbox{\scriptsize #1}}
\begin{document}

\tighten

\title{
Evidence for Braggoriton Excitations in Opal Photonic Crystals 
Infiltrated with Highly Polarizable Dyes
}

\author{
N. Eradat$^*$, A. Y. Sivachenko, M. E. Raikh and Z. V. Vardeny}

\address{
Department of Physics, University of Utah, salt Lake City, Utah 84112
}

\author{A. A. Zakhidov and R. H. Baughman}
\address{Honeywell Int., Research and Technology Center, Morristown, New Jersey 07962}

\maketitle

\begin{abstract}

We studied angle-dependent reflectivity spectra of opal photonic crystals infiltrated with 
cyanine dyes, which are highly polarizable media with very large Rabi frequency. We 
show that when resonance conditions between the exciton-polariton of the infiltrated dye 
and Bragg frequencies exist, then the Bragg stop band decomposes into two reflectivity 
bands with a semi-transparent spectral range in between that is due to light propagation 
inside the gap caused by the existence of braggoriton excitations. These novel excitations 
result from the interplay interaction between the Bragg gap with spatial modulation origin 
and the polariton gap due to the excitons, and may lead to optical communication traffic 
inside the gap of photonic crystals via channel waveguiding.

\end{abstract}

\vskip6mm

Photonic crystals (PC)\cite{ref1} and in particular PC with a complete photonic 
band gap (PBG)\cite{ref3} have recently attracted much attention due to their rich 
physics and possible 
applications such as lasers, optical communication and tunable, high reflectivity 
devices\cite{ref4,ref5}. In these systems the dielectric function is periodically 
modulated and consequently light diffraction effects dominate their optical 
properties\cite{ref6}. When Bragg diffraction 
conditions are met then light scattering is very strong so that a reflectivity 
plateau occurs 
within a certain frequency interval, $\Delta\omega_{\scr{B}}$ 
around the Bragg resonance, $\omega_{\scr{B}}$, where light 
propagation is inhibited (see Figs.1(a) and 1(b)). Among the main possible applications 
of PC is the fabrication of low threshold lasers\cite{ref7,ref9}, 
since the photonic density of states 
can be enhanced for frequencies close to the PBG\cite{ref1}. 
Photonic crystals that contain 
few planned defects are most suitable for laser action, since defects cause localized 
intragap photonic states for which the PC sample acts as a resonator with a very high 
quality factor, $Q$\cite{ref7}. However, photon propagation traffic in and out from 
the defect-resonator has not been well developed for three-dimensional (3D) PC\cite{ref9}. 
In fact there 
has been only little experimental progress on light propagation inside the PBG and 
mostly for line defects in 2D PC\cite{ref9,ref10,ref11,ref12,ref13}.
 
In this work we demonstrate a way by which light can, in fact communicate with a 
high $Q$ laser resonator with frequency inside the gap of a 3D PC. In particular we show 
that intragap light propagation in 3D PC may be controlled by wave-guiding due to novel 
excitations, dubbed here braggoritons\cite{ref14}, which are present inside the PBG when the PC is 
infiltrated with a {\em highly polarizable} medium having strong coupling to
light. PC's infiltrated with polarizable media having {\em weak} light coupling were
treated in Ref.~\onlinecite{ref15}. The braggoriton excitations are 
formed
through the interaction between the Bragg gap due to spatial modulation and polariton 
gap due to the excitons, when their characteristic frequencies ($\omega_{\scr{B}}$ and 
$\omega_{\scr{T}}$, respectively) 
are close to each other. As a result, a semi-transparent spectral region is formed 
inside the 
gap, which splits the Bragg plateau in reflectivity into two peaks such that intragap light 
propagation becomes possible for frequencies in between the peaks. We demonstrate the 
existence of the braggoriton excitations in a 3D opal PC infiltrated with a cyanine dye 
that has very large Rabi frequency\cite{ref16,ref17} and hence is highly polarizable. 
In contrast to 
recent experiments\cite{ref17} where the infiltrated dye and 
Bragg reflectors do not spatially 
overlap and thus the braggoriton excitations are not formed, in our case the polarizable 
medium and the Bragg reflectors coexist, giving rise to the novel excitations. We show 
that the angle-dependent reflectivity spectra of the infiltrated opal contain a split Bragg-
polariton stop band with a semi-transparent band in between where braggoriton 
excitations can propagate for impinging angles, $\theta$ at which the unperturbed Bragg stop 
band, $\omega_{\scr{B}}(\theta)$ is close to $\omega_{\scr{T}}$. The correlation of the 
two reflectivity bands to each other is 
further demonstrated when, due to a strong light bleaching effect of the cyanine 
molecules the two reflectivity bands collapse together as the cyanine molecules are 
progressively bleached with the illumination time. 

The synthetic opal single crystals used here were cut from a polycrystalline sample 
obtained by slow sedimentation of a colloidal suspension of silica (SiO$_2$) spheres (with a 
mean diameter $D\approx300\;$nm and dispersion in $D$ of about 4\%\cite{ref19}). 
Since the as-grown 
opals are weakly bound, sintering at 750$^{\circ}$C was used to achieve robust mechanical 
properties. This sintering process provides inter-sphere necks joining neighboring silica 
spheres in a face-centered-cubic (fcc) structure of the opal. Weak iridescence in air 
results from Bragg diffraction off $(hkl)$ crystal planes, which produce gaps in the 
reflectivity spectrum at $\omega_{\scr{hkl}}$\cite{ref20}. However due to the 
relatively small contrast between 
the dielectric constant, $\varepsilon$ of the silica and air, the opal PC do 
not possess a complete 
PBG; instead the various $\omega_{\scr{hkl}}$ 
\parbox{8.9cm}{
\psfig{file=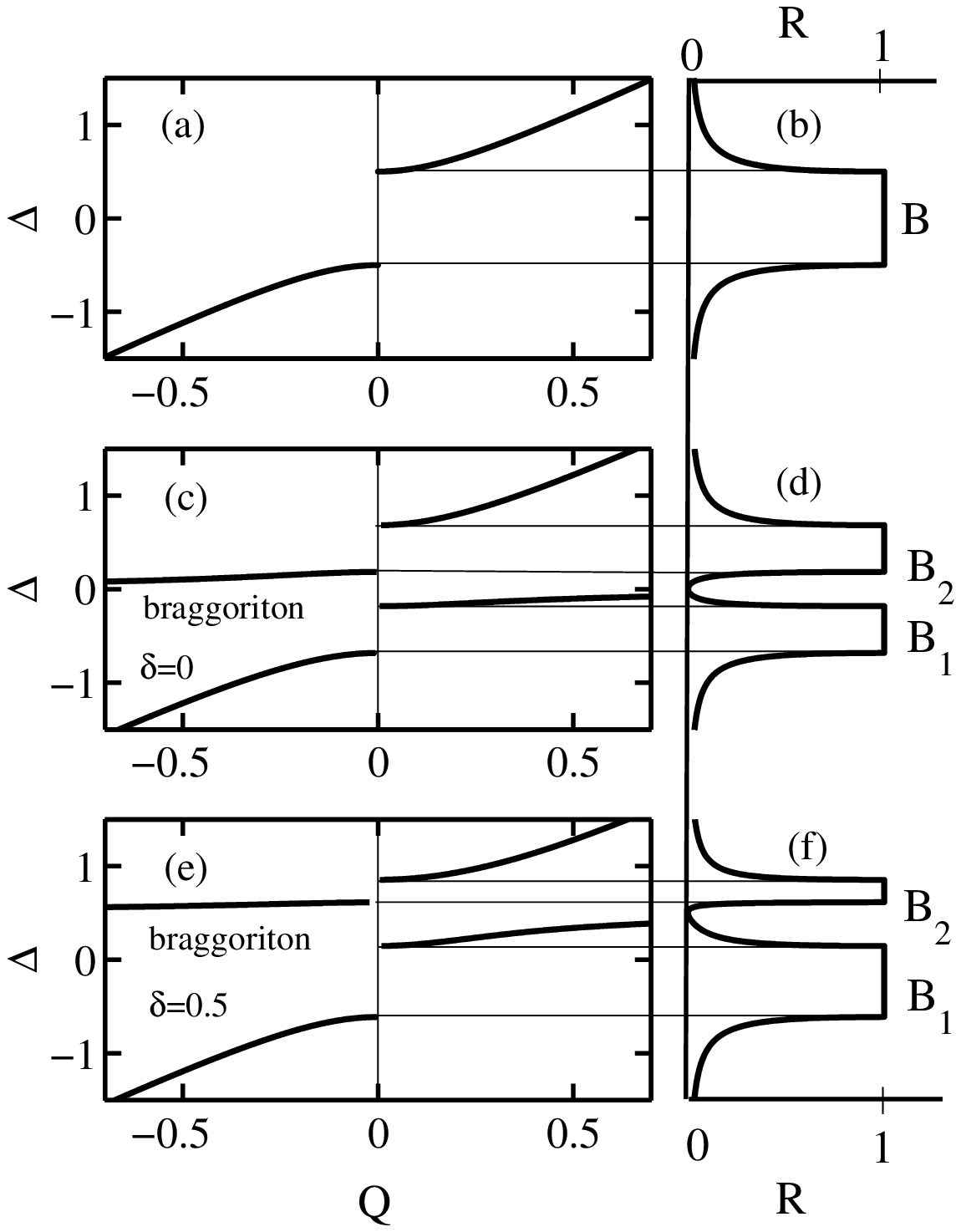,width=8.3cm}

\vskip1mm

{\small {\bf Fig. 1.} The dispersion relations [$\Delta(Q)$, where $\Delta$ and $Q$ 
are normalized frequency and 
wave-vector, respectively (see text)] and reflectivity spectra [$R(\Delta)$] 
of an uninfiltrated (a) 
and (b), and infiltrated (c)-(f) PC with a highly polarizable dye, for wave-vector $Q$ close 
to the Brillouin edge ($Q = 0$). For the infiltrated opal the normalized detuning between the 
Bragg and exciton/polariton gaps is $\delta =0$ for (c) and (d) and $\delta = 0.5$ 
for (e) and (f), 
whereas the normalized coupling parameter is $\alpha =0.5$ in both cases. $B$ is the Bragg stop 
band in reflectivity, whereas $B_1$ and $B_2$ are the split stop bands due to Bragg-like and 
polariton-like gap, respectively.
}

\vskip3mm

}
\hskip5mm\parbox{8.3cm}{
\noindent
do not overlap in the spectrum. 
In general the most 
pronounced stop band in synthetic opals occurs along the fast growing [111] 
direction\cite{ref21}, and therefore we conduct our angle-dependence reflectivity 
measurements off the (111) planes.

Along the preferred [111] (or $z$) direction we may simplify the treatment of light 
propagation by considering the opal structure as a one-dimension system. Thus along the 
z-direction the dielectric constant of the uninfiltrated opal can be written as: 
$\varepsilon(z) = \varepsilon_0 + 
\delta\varepsilon\cos(Kz)$, where $\delta\varepsilon$ ($\ll\varepsilon_0$) 
is the modulation amplitude, $\varepsilon_0$ is the average opal 
dielectric constant, $K = 2\pi/d$ is the modulation wave-vector, and d is the (111) 
interplane 
distance given by: $d = D\sqrt{2/3}$. Assuming that the electromagnetic field propagates 
along 
$z$ and neglecting polarization effect, it is easy to show\cite{ref4} that a gap, 
$2\Delta\omega_{\scr{B}}$ is formed in 
the photon dispersion relations at a frequency 
$\omega_{\scr{B}} = cK/(2\sqrt{\varepsilon_0})$ for light with wave-vector k 
close to the Brillouin zone at $K/2$, where 
$\Delta\omega_{\scr{B}} = \omega_{\scr{B}}\delta\varepsilon/(2\varepsilon_0)$ 
(see Fig.1(a)). In addition, 
the two frequency branches at wave-vector $q$ away from $K/2$ (where $q = k-K/2$) acquire 
the form
\begin{equation}
\omega^{(1,2)} = \omega_{\scr{B}}\pm \sqrt{\left(\frac{2\omega_{\scr{B}}}{K}\right)^2
q^2+\Delta\omega_{\scr{B}}^2}.
\end{equation}

These dispersion relations are plotted in Fig. 1(a); in Fig. 1(b) we plot the resulting 
calculated Bragg stop band in the reflectivity spectrum. It is worth noting that when light 
propagates at an angle $\theta$ respect to $z$ (or [111]), then the Bragg stop band 
blue shifts according to the relation\cite{ref20}
\begin{equation}
\omega_{\scr{B}}(\theta) = \omega_{\scr{B}}(0)\sqrt{\frac{\varepsilon_0}
{\varepsilon_0-
\varepsilon_1\sin^2\theta}}
\end{equation}
where $\varepsilon_1$ is the dielectric constant of the surrounding medium. \hfill
Our \hfill opal \hfill photonic \hfill crystal \hfill 
was \hfill composed \hfill of silica sphere having 
\hfill a mean \hfill diameter $D$ \hfill of \hfill about
}
300$\;$nm 
with
$\varepsilon_0 \approx1.55$\cite{ref19}; these parameters 
determine the 
Bragg stop band in air to be close to 660 nm at $\theta=0$\cite{ref22}. 
This stop band, however could be easily shifted with $\theta$ to about 560 nm for 
$\theta= 45^{\circ}$ (Eq. (2))\cite{ref22}.

The photon dispersion relations and the reflectivity spectrum drastically 
change\cite{ref14}
when a polarizable medium having a dielectric constant 
$\varepsilon(\omega) = \varepsilon_{\infty}[1-\omega_{\scr{LT}}/(\omega-\omega_{\scr{T}})]$,
where  $\omega_{\scr{T}}$ is the transverse frequency, 
$\omega_{\scr{LT}}$ is the longitudinal-transverse frequency splitting 
[with a Rabi frequency splitting 
$\Omega_{\scr{P}}= \sqrt{\omega_{\scr{T}}\omega_{\scr{LT}}/2}$, and 
$\varepsilon_{\infty}$ is the high frequency dielectric 
constant, is infiltrated into the opal. In this case if the frequency detuning, given by the 
normalized detuning parameter,
$\delta = (\omega_{\scr{T}}-\omega_{\scr{B}})/(2\Delta\omega_{\scr{B}})$ is not too large 
(i.e. $\delta<1$) then the light 
dispersion relations near $\omega_{\scr{B}}$ contain four branches that are 
given\cite{ref14} in a concise form by: 
\begin{equation}
Q = \sqrt{\left[\Delta-\frac{\alpha^2}{\Delta-\delta}\right]^2-\frac{1}{4}},
\end{equation}
where $Q$ is the normalized wave-vector, $Q = (\omega_{\scr{B}}/K\Delta\omega_{\scr{B}})q$,
$\alpha$ is the Bragg-polariton 
coupling constant given by $\alpha = \Omega_{\scr{P}}/2\Delta\omega_{\scr{B}}$, and 
$\Delta$ is the normalized frequency given by
$\Delta = (\omega-\omega_{\scr{B}})/(2\Delta\omega_{\scr{B}})$. 
These dispersion relations are plotted in Fig. 1(c) for a detuning $\delta = 0$; 
the resulting calculated reflectivity spectrum is shown 
in Fig.1(d). It is seen that the 
interaction between the Bragg gap and the exciton/polariton gap leads to light 
propagation inside the photonic gap via novel type excitations, dubbed here braggoritons. 
This is evident in the dispersion relations (Fig. 1(c)) as well as in the transparent 
spectral  range that is formed in between the two reflectivity plateaus in Fig.1(d). 
\noindent
\parbox{8.9cm}{

\psfig{file=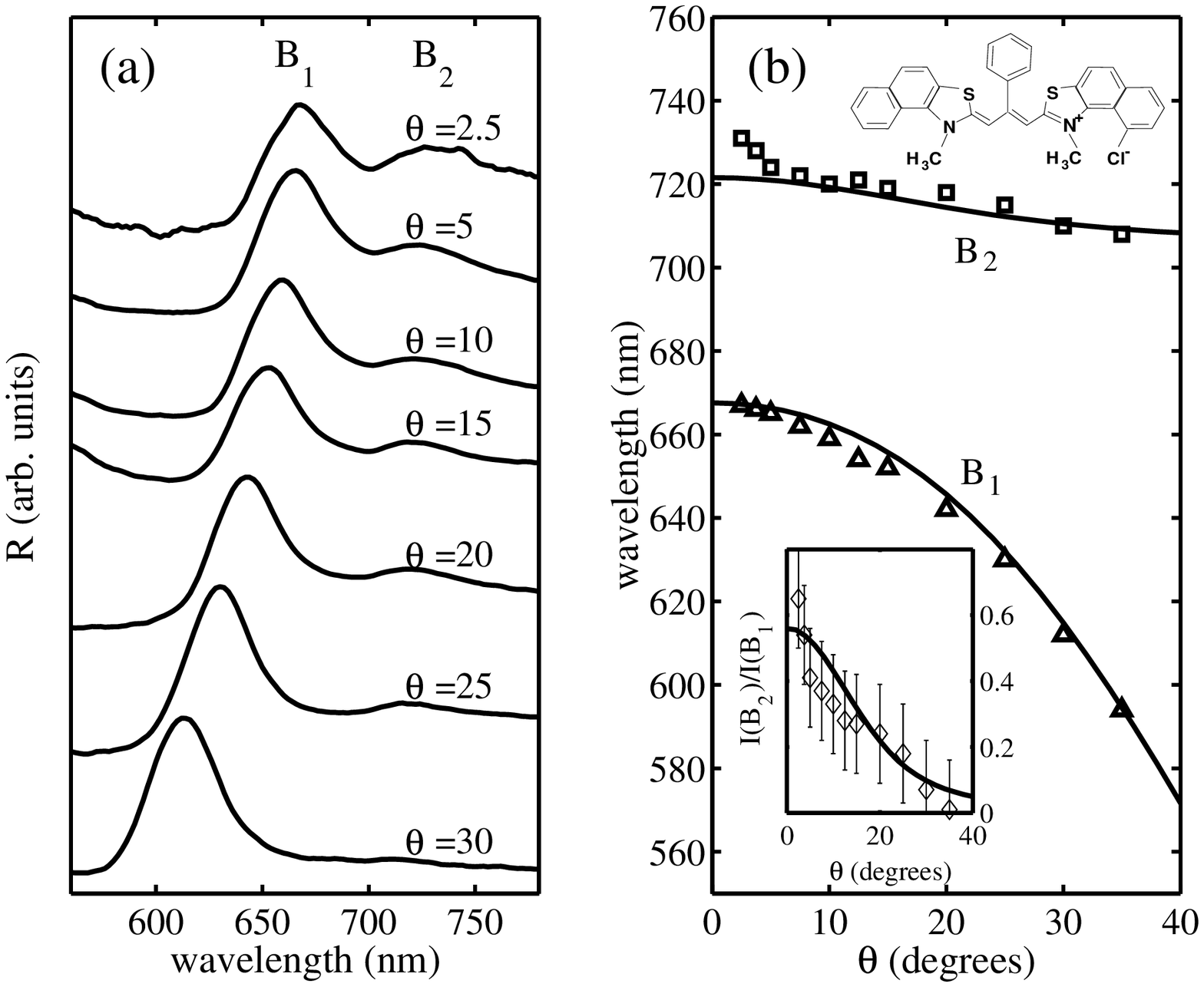,width=8.3cm}

\vskip1mm

{\small {\bf Fig. 2.} (a) The angular dependent reflectivity spectra of an opal PC infiltrated with J-
aggregates of NK-2567 cyanine that show the evolution of the two reflectivity bands $B_1$ 
(Bragg-like) and $B_2$ (polariton-like) with the light impinging angle, $\theta$.  
The frequency 
dispersion and relative strengths of $B_1$ and $B_2$ bands with $\theta$ 
are summarized in (b) and (b) 
inset, respectively.
}

\vskip2mm}
\hskip5mm\parbox{8.3cm}{

\hskip3mm For a detuning $\delta\neq 0$ the photon dispersion relations lose their 
symmetry\cite{ref14} (Fig. 
1(e)) and this can be also 
seen in the reflectivity spectrum (Fig. 1(f)), where the 
reflectivity peak due to the polariton-like branch in the 
dispersion relations 
($B_2$ in Fig.1(e)) diminishes with increasing $\delta$. By changing the light 
impinging angle $\theta$ respect to the 
[111] direction, the detuning $\delta$ between the Bragg and polariton gaps may be easily 
changed in opals since $\omega_{\scr{B}}(\theta)$ changes with $\theta$ according to 
Eq. (2) above, whereas $\omega_{\scr{T}}$ 
remains unaffected when $\theta$ is varied\cite{ref23}. We use this mechanism 
of 
changing $\delta$ in the 
infiltrated opals 
for measuring the dispersion of the braggoriton excitations with~$\theta$.

\hskip3mmFor these measurements we infiltrated a large single crystal opal with the cyanine 
dye NK-2567, or 2,2'-dimethyl-8-phenyl-5, 6, 5', 6'-dibenzothiacarbocynine chloride, of 
which chemical diagram is shown in Fig.2(b), inset. In chloroform solution these cyanine 
molecules weakly absorb in the yellow part of the spectrum (at approx. 600 nm) and have 
a moderately strong photoluminescence band in the red\cite{ref17}. However when a thick 
chloroform solution of NK-2567 is casted into films on glass substrates, then J-aggregates 
are readily formed from the cyanine molecules. The cyanine J-aggregates are 
characterized by a red shifted 
}
absorption (or reflectivity) band that peaks at 
$\omega_{\scr{T}}\approx 700\;$nm; 
simultaneously, the photoluminescence band
greatly diminishes so that clean reflectivity 
measurements could be completed with little or no interference from the emission 
band\cite{ref17}. The aggregate formation is not as straightforward when the cyanine 
molecules are 
infiltrated inside the opal, since the opal voids are not aligned properly to promote 
aggregation. However repeatably pulling the opal sample from the thick solution 
provided better aligning to form the J-aggregates inside the opal\cite{ref22}. 
In addition, the 
aggregated cyanine molecules on the opal surface were carefully washed-out so that an 
extra reflectivity peak in the spectrum\cite{ref23}, which is due to the uncoupled 
excitons/polaritons, does not form to further complicate the spectrum. The opal sample 
was embarked on a $\theta$--$2\theta$ homemade goniometer, where an incandescent, 
well-collimated light beam was directed at an angle $\theta$ respect to the [111] 
direction of the opal. 
The reflected beam at $2\theta$ was dispersed by a monochromator (0.25 m) and its intensity 
measured by a Si photodiode and a lockin amplifier\cite{ref22}. At $\theta  = 0$ 
the uninfiltrated 
opal showed a [111] Bragg stop band in the reflectivity spectrum at 
$\omega_{\scr{B}}(0)\approx 660\;$nm; this band 
red-shifted when infiltrated with the cyanine aggregates due to the increase in 
$\varepsilon_0$ upon 
infiltration so that the new $\omega_{\scr{B}}(0)\approx\omega_{\scr{T}}\approx 700\;$nm.  

Fig. 2(a) shows the angle-dependent reflectivity spectra of the cyanine-infiltrated 
opal for $\theta$ ranging from 2.5$^{\circ}$ to 30$^{\circ}$. At large $\theta$ when 
$\omega_{\scr{B}}(\theta) > \omega_{\scr{T}}$ the Bragg/polariton 
interaction is small. Then due to the large detuning $\delta$, only one reflectivity 
band ($B_1$), 
which is the Bragg stop band, dominates the spectrum. However as $\theta$ decreases, 
$\omega_{\scr{B}}(\theta)$ 
red shifts towards $\omega_{\scr{T}}$ so that the detuning $\delta$ decreases and 
consequently the 
Bragg/polariton interaction increases. It is then apparent that another band ($B_2$) 
is formed 
at small $\theta$, with increasing relative intensity at smaller $\theta$ 
(Fig.2(b), inset). At the 
same time the dispersion of peak $B_1$ slows down, whereas the dispersion of peak 
$B_2$ 
increases (Fig. 2(b)). This behavior is exactly as expected from the braggoriton model 
above if there is strong coupling between the Bragg and polariton gaps when they 
are close to each other\cite{ref14}. The reason that a deeper dip in between the split reflectivity 
bands is not observed in the reflectivity spectra at small 
$\theta$ is the unavoidable disorder that 
exists in the opal sample, which tends to smear out sharp spectral features\cite{ref24}. 

We used the braggoriton model\cite{ref14} to fit the experimental data. For this fit we 
calculated the Bragg-like ($B_1$) and polariton-like ($B_2$) reflectivity peaks in the 
angle-dependent spectra to derive the peaks frequency and relative intensity at each 
$\theta$. The 
calculated frequencies and relative intensities versus $\theta$ are shown in Figs. 2(b) 
and 2(b) 
inset, respectively and compared with the data. The good agreement between theory and 
experiment seen in Figs. 2(b) and 2(b) inset was achieved using Eq.~(1) with the 
following free parameters: $\omega_{\scr{B}} = 683\;$nm, $\Delta\omega_{\scr{B}}= 53\;$meV 
(or 20 nm), which are slightly shifted compared to the values of uninfiltrated opal;
$\omega_{\scr{T}} = 705\;$nm, which is directly determined from the reflectivity peak
in the aggregated films outside the opal; and $\varepsilon_0 =1.45$: this results in a 
detuning parameter $\delta=0.55$ at $\theta = 0$. In addition, we used a Rabi 
frequency splitting 
\noindent
\hbox{\parbox{8.9cm}{
\psfig{file=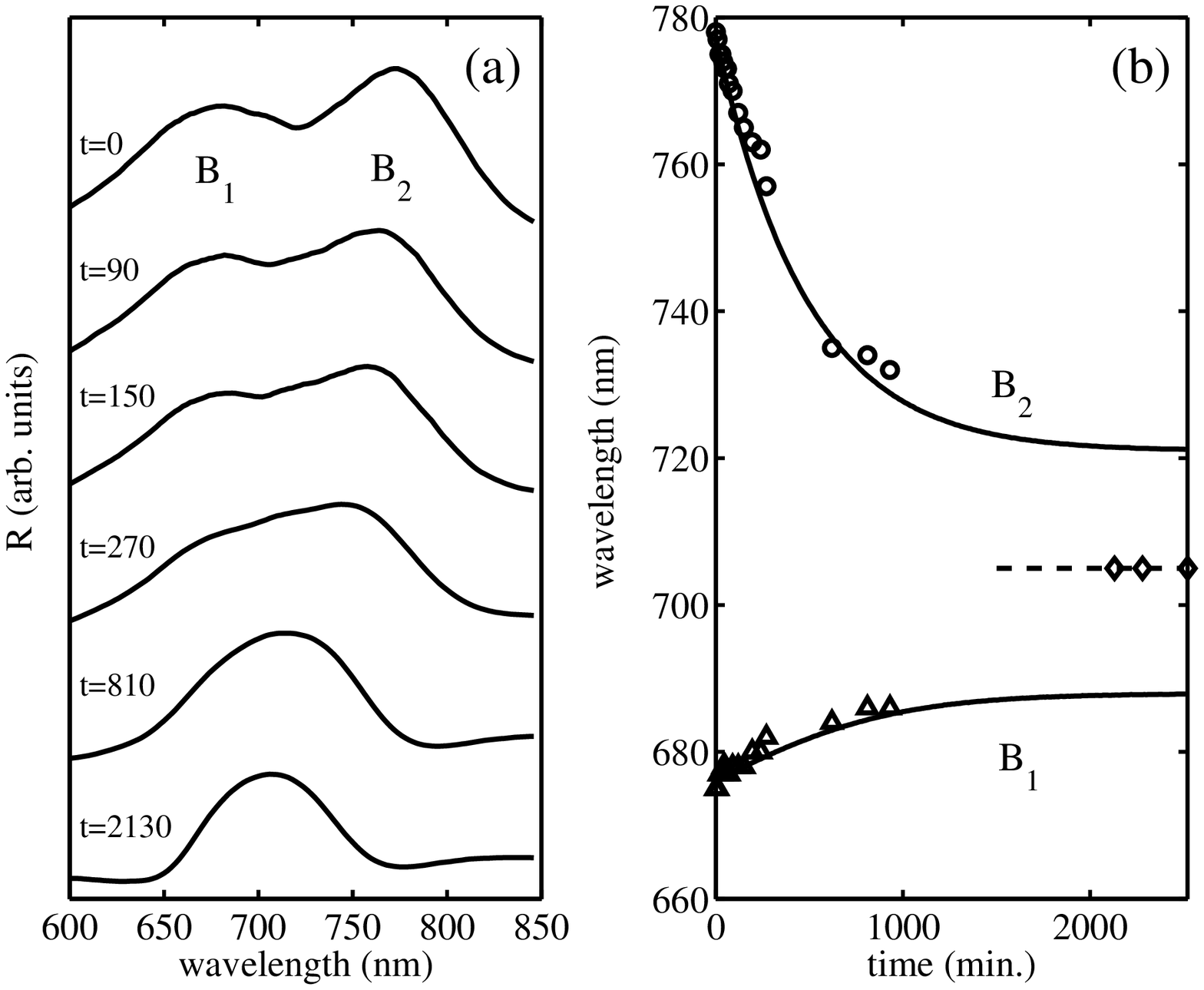,width=8.3cm}

\vskip1mm

{\small {\bf Fig. 3.} (a) The dependence of the reflectivity spectrum of NK-2567 cyanine dye 
infiltrated into opal measured at $\theta = 0$, on the illuminating time, $t$ 
in minutes using a 
strong incandescent light to bleach the dye molecules. (b) The frequency dependence of 
$B_1$ and $B_2$ bands in (a) on the illumination time, $t$; triangles are for the 
$B_1$-branch and 
circles are for the $B_2$-branch; the diamonds symbols are for the collapsed band at large $t$, 
when $B_1$ and $B_2$ cannot be separated any more.     
}

\vskip8mm

}
\hskip6mm\parbox{8.3cm}{
$\Omega{\scr{P}} = 66\;$meV for the actual NK-2567 J-aggregates 
density that  
infiltrated the opal crystal. From $\Omega_{\scr{P}}$ and $\Delta\omega_{\scr{B}}$ 
above we calculate a relatively strong 
Bragg-polariton coupling constant $\alpha = 0.62$. Taking into consideration that the Rabi 
frequency splitting, $\Omega_{\scr{P}}$ depends on the density of infiltrated 
aggregates into the opal, 
which may change 
from one measurement set to the other, then the obtained $\Omega_{\scr{P}}$
of 66 meV is in good agreement with that deduced previously from reflectivity measurements 
of the same cyanine dye in a tilted micro-cavity, where 
$\Omega_{\scr{P}}$ of 80 meV was obtained\cite{ref17}.

\hskip3mm We found that the coupling parameter $\alpha$ can be readily 
varied in the infiltrated opal 
by changing the density of the infiltrated cyanine molecules\cite{ref22}. 
This occurs since the 
cyanine Rabi frequency splitting, $\Omega_{\scr{P}}$ is proportional to 
$\sqrt{\omega_{\scr{LT}}}$, with $\omega_{\scr{LT}}$ being in turn proportional 
to the molecules density, $N$, where $\omega_{\scr{LT}} = 2\pi Ne^{*2}/m\omega_{\scr{T}}$. 
By changing $N$ it is actually 
possible to change the reflectivity spectrum of the infiltrated opal at a fixed 
$\theta$, since the 
changing coupling parameter $\alpha$ determines the splitting between the 
Bragg and polariton 
reflectivity bands. We use this relation to directly prove the correlation that exists 
between peaks $B_1$ and $B_2$ in the reflectivity spectrum. 
Fig. 3(a) shows the evolution of the 
infiltrated opal reflectivity spectrum at $\theta = 0$ 
with the illumination time of a strong 
incandescent light source, which is used here to bleach~the
}}
cyanine molecules. It is seen 
that peaks $B_1$ and $B_2$ progressively approach 
each other with the illumination 
bleaching time, $t$ as summarized in Fig. 3(b). 
At long times both peaks collapse into a 
single band at
about 700 nm (Fig. 3(a)), which is in agreement with 
the frequencies $\omega_{\scr{B}}$ and 
$\omega_{\scr{T}}$ used 
above to fit the angular dependence of the reflectivity spectra in Fig.~2. The evolution of 
the peak frequencies of both branches with the illumination time can be very well 
described (Fig. 3(b)) by a single parameter $\alpha(t)$, or $N(t)$, where 
$N(t)$ decays exponentially, 
$N(t) = N(0)exp(-t/\tau)$ with a time constant, $\tau = 425\;$minutes that characterizes the 
bleaching process of the cyanine molecules with the illumination time, and 
$\alpha(0) = 0.75$, which is slightly larger than the film in Fig.~2. 
The excellent fit shown in Fig. 3(b) for the dispersion of both branches with time was 
achieved by taking into consideration the change in the dielectric constant $\varepsilon_0$ with time, 
where $\epsilon_0(t) = 1.5 + 0.16 exp(-t/\tau)$; this affects both the 
Bragg frequency and its 
width so that both parameters are time-dependent with $\lambda_{\scr{B}}(0) = 746\;$nm 
and $\Delta\omega_{\scr{B}}(0) = 30\;$nm; $\omega_{\scr{T}}$ was already determined from Fig.~2.
The good agreement between experiment and theory unambiguously shows that the 
Bragg-like and polariton-like bands are correlated to each other in the infiltrated opal and
that the braggoriton model is correct. Furthermore the quantitative fits in both Figs.~2 and~3
may be taken as evidence for the existence of braggoriton light excitations in opals infiltrated
with highly polarizable media.

The infiltrated polarizable medium in opal may serve as a channel waveguide for in and 
out light traffic from the PC, for frequencies inside the gap. In general, even in the 
absence of the Bragg grating effect a channel structure having a dielectric constant 
$\varepsilon_1$ that 
is larger than $\varepsilon_2$ 
of the surrounding always supports a waveguide mode along the channel 
direction, $z$ due to the total internal reflection at the interfaces. This mode is 
localized in 
the lateral direction, $x$ with an attenuation length, $L_x$, which is determined by 
the ratio 
$\varepsilon_1/\varepsilon_2$. In the presence of Bragg grating along $z$ 
and for frequencies inside the PBG, 
however a waveguide mode exists only if $L_x<L_x^0 = (L_{\scr{B}}/K)^{1/2}$, where 
$L_{\scr{B}}$ [$= 2\omega_{\scr{B}}/(K\Delta\Omega_{\scr{B}})$] 
is the Bragg attenuation length for $\omega =\omega_{\scr{B}}$. 
If the channel is filled with a polarizable 
medium then for $\omega\approx\omega_{\scr{T}}$, $\varepsilon_1$ 
is large and consequently $L_x$ substantially decreases. At some 
frequency, $\omega_0 < \omega_{\scr{T}}$, however $L_x$ becomes so 
small that a waveguide mode is formed in 
the structure along~$z$. 

       We calculated $L_x$ for a channel with lateral dimension $d <<L_x^0$. 
We found in the case $\omega_{\scr{T}}\approx\omega_{\scr{B}}$, 
$\omega < \omega_{\scr{T}}$ and $(\omega - \omega_{\scr{B}}) << \omega_{\scr{B}}$, that
\begin{equation}
 L_x \approx \frac{8(\omega_{\scr{T}} - \omega)}{\omega_{\scr{LT}}K^2d}.  
\end{equation}
Using the condition for the existence of a wave-guide mode, namely $L_x < L_x^0$ we find 
the waveguide/braggoriton frequency in the PBG to be: 
$\omega>\omega_0 = \omega_{\scr{T}} - (1/8)\omega_{\scr{LT}}Kd
\sqrt{2\omega_{\scr{B}}/\Delta\omega_{\scr{B}}}$. 
We note that $\omega_0 < \omega_{\scr{T}}$ and therefore the waveguide-braggoriton 
mode can propagate for relatively long distances along $z$ with little 
attenuation due to absorption. 
                 
       In conclusion, we showed that a highly polarizable medium that is infiltrated into a 
PC can induce a transparent spectral region inside the gap when the coupling, or Rabi 
frequency is strong and the unperturbed polariton frequency is close to the gap frequency. 
This is caused due to the strong interaction between the Bragg and exciton-polariton gaps 
that results in the existence of intragap braggoriton excitations, which promote light 
propagation inside the gap. Although the braggoriton-induced transparency was 
demonstrated for cyanine dye aggregates infiltrated in an opal PC that does not possess a 
complete PBG, we expect that our conclusions would also hold in case of a PC with a 
complete PBG. The induced intragap transparency may then serve to direct light traffic in 
and out of a defect laser resonator inside the PC via a braggoriton type waveguide. 

This work was supported in part by the Army Research Office grant DAAD 19-00-1-0406, 
the Petroleum Research Fund grant ACS-PRF 34302-AC6, and DARPA grant 
DAA-99-J-036.

\end{document}